\documentstyle[12pt]{article}
\input epsf.sty
\newdimen\psfigsize
\def\psfigure#1 #2 #3 #4 #5{
    \begin{figure}[tbp]
    \vbox{
    \null\hskip#2\epsfxsize=#1 \epsfbox[0 0 4096 4096]{#4}
    \vskip 10truept
    \caption {#5 \label{#3}}
    \vskip 0.1truein plus0.2truein}
    \end{figure}
}
\def\pspagefigure#1 #2 #3 #4 #5{
    \begin{figure}[p]
    \vbox{
    \null\hskip#2\epsfxsize=#1 \epsfbox[0 0 4096 4096]{#4}
    \vskip 10truept
    \caption {#5 \label{#3}}
    \vskip 0.1truein plus0.2truein}
    \end{figure}
}
\def\psoddfigure#1 #2 #3 #4 #5 #6{
    \begin{figure}[tbhp]
    \vbox{
    \null\hskip#3\epsfxsize=#1 \epsfbox[0 0 4096 4096]{#5}
    \vskip -#1 \vskip #2 \vskip 10truept
    \vskip 10truept
    \caption {#6 \label{#4}}
    \vskip 0.1truein plus0.2truein}
    \end{figure}
}
\def\figurespace#1 #2 #3 #4 {
    \begin{figure}[tbhp]
    \vbox{
    \psfigsize=#1truein
    \vskip \psfigsize
    \vskip 10truept
    \caption {#4 \label{#3}}
    \vskip 0.1truein plus0.2truein}
    \end{figure}
}
\def\gnufigure#1 #2 #3 #4 #5 #6{
    \begin{figure}[tbhp]
    \vbox{
    \null\hskip#3\epsfxsize=#1 \epsfbox{#5}
    \vskip -#1 \vskip #2 \vskip 10truept
    \vskip 10truept
    \hbox{\null\hskip 1.0in \parbox[t]{4.5in}{ \caption {#6 \label{#4}} } }
    \vskip 0.1truein plus0.2truein}
    \end{figure}
}

\def\etal{{\it et al.\ }}

\def\Dslash{\mathop{\not\!\! D}}

\def\LP{\left(}		
\def\RP{\right)}	
\def\LB{\left\{}	
\def\RB{\right\}}	

\def\BE{\begin{equation}}
\def\EE{\end{equation}}
\def\BEA{\begin{eqnarray}}
\def\EEA{\end{eqnarray}}

\begin{document}

\begin{titlepage}
\baselineskip=16pt
\rightline{\bf hep-lat/9609036}
\rightline{AZPH-TH/96-22}
\baselineskip=20pt plus 1pt
\vskip 1.5cm

\centerline{\Large \bf Improving flavor symmetry in the}
\centerline{\Large \bf Kogut-Susskind hadron spectrum}
\bigskip
\centerline{\bf Tom~Blum }
\centerline{\it
Department of Physics, Brookhaven National Lab, Upton, NY 11973, USA
}
\centerline{\bf Carleton~DeTar}
\centerline{\it
Physics Department, University of Utah, Salt Lake City, UT 84112, USA
}
\centerline{\bf Steven~Gottlieb, Kari~Rummukainen }
\centerline{\it
Department of Physics, Indiana University, Bloomington, IN 47405, USA
}
\centerline{\bf Urs~M.~Heller }
\centerline{\it
SCRI, Florida State University, Tallahassee, FL 32306-4052, USA
}
\centerline{\bf James~E.~Hetrick, Doug~Toussaint }
\centerline{\it
Department of Physics, University of Arizona, Tucson, AZ 85721, USA
}
\centerline{\bf R.L.~Sugar }
\centerline{\it
Department of Physics, University of California, Santa Barbara, CA 93106, USA
}
\centerline{\bf Matthew~Wingate }
\centerline{\it
Physics Department, University of Colorado, Boulder, CO 80309, USA
}

\narrower
We study the effect of modifying the coupling of Kogut-Susskind quarks
to the gauge field by replacing the link matrix in the quark action by
a ``fat link'', or sum of link plus three-link paths.  Flavor symmetry
breaking, determined by the mass difference between the Goldstone and
non-Goldstone local pions, is reduced by approximately a factor of
two by this modification.
\end{titlepage}


\vskip0.25in\centerline{Introduction}

One of the major technical challenges facing lattice QCD is the
extraction of continuum physics from lattices with few enough lattice
sites
that computations are feasible.  Recently there has been progress in
developing ``improved actions'', which give better approximations to
continuum physics with large lattice spacings\cite{ACTIONREVIEW}.
Starting points for improving the gauge field action include 
cancellation of lattice artifacts in an expansion in powers of $a$
(Symanzik actions)\cite{PUREG,ALFORD}
or renormalization group ideas (perfect actions)\cite{PERFECT}.  Improved
fermion actions which improve the quark dispersion relation have been
introduced for Wilson quarks (``clover'' action)\cite{CLOVER} and for
Kogut-Susskind quarks\cite{NAIK}. In particular, the Naik action for
Kogut-Susskind quarks adds a third-nearest-neighbor term to remove an
unwanted term of order $a^2$, relative to the desired term, from the
dispersion relation.  This produces improvements such as energy and
baryon number susceptibilities at high temperature that quickly approach
the free field (Stefan-Boltzmann) limits, and has been applied to
QCD thermodynamics\cite{KARSCHLATTICE96}.
Another of the major problems with Kogut-Susskind quarks is the breaking
of flavor symmetry.  In particular, only one of the pions is a true
Goldstone boson at nonzero lattice spacing, and at the lattice spacings
where simulations are done the differences in the pion masses are
large.  Preliminary results suggest that improving the gauge action
helps this problem, but the Naik improvement of the fermion action does
not\cite{MILCIMPROVED}.  This is not surprising, since the motivation
for the Naik action is the improvement of the free quark dispersion
relation rather than the flavor symmetry.

Here we explore a simple modification of the Kogut-Susskind fermion
action.  In particular, we replace the link matrices in the fermion
matrix by weighted sums of the simple link plus three link paths, or
``staples'', connecting the points.  This is a simple modification that
is consistent with gauge invariance and the hypercube symmetries of the
lattice.
Modifications such as this are likely to
arise in any renormalization group motivated improved fermion action ---
this is just the simplest possible addition.  In simulations with
dynamical fermions, simplicity will be important since the computation
of the fermion force in the molecular dynamics integration could become
impossibly complicated.  Of course, this modification of the quarks'
parallel transport can, and probably should, be used in concert with
improvements in the gauge field action and the Naik (third nearest
neighbor) improvement to the quark dispersion relation.

In this paper we report on a study of the quenched meson spectrum using
the ``fat link'' quark action.  Quenched spectrum calculations are easy
to do, especially when stored configurations are available, and there
are many results in the literature using the conventional fermion action
for comparison.  We measure masses of the ``local'' mesons, which
include the Goldstone pion and one non-Goldstone pion, the ``SC'' pion,
or ``$\pi_2$''\cite{EDINBURGH}.
The splitting between these two pions is a good indicator of flavor
symmetry breaking, since the calculations that have measured other pion
masses find that all of the non-Goldstone pions are nearly
degenerate\cite{ALLPIONMASSES}.

\vskip0.25in\centerline{Formalism}

We modify the standard Kogut-Susskind Dirac operator by replacing each
link matrix $U_\mu(x)$ by a ``fat link''
\BE
U_{fat,\mu}(x) = \frac{ U_\mu(x) + w\sum_{\nu \ne \mu} \left(
U_\nu(x) U_\mu(x+\hat\nu) U_\nu^\dagger (x+\hat\mu) +
U_\nu^\dagger(x-\hat\nu) U_\mu(x-\hat\nu) U_\nu(x-\hat\nu+\hat\mu)
\right) }
{1 +6w}
\EE
where $w$ is an adjustable weight for the staples.

Symbolically, this is just:

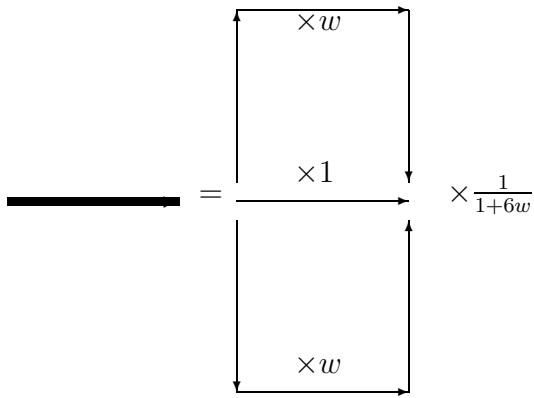
\setlength{\unitlength}{1.0in}
\begin{picture}(6.0,2.3)(-3.0,-1.15)	

\thicklines
\linethickness{1.0mm}
\put(-1.0,0.0){\vector(1,0){0.9}}
\put(0.0,0.0){=}
\thinlines
\put(0.5,0.1){$\times 1$}
\put(0.5,0.9){$\times w$}
\put(0.5,-0.9){$\times w$}
\put(1.3,0.0){$\times\frac{1}{1+6w}$}
\put(0.2,0.0){\vector(1,0){0.9}}
\put(0.2,0.1){\vector(0,1){0.9}}
\put(0.2,1.0){\vector(1,0){0.9}}
\put(1.1,1.0){\vector(0,-1){0.9}}
\put(0.2,-0.1){\vector(0,-1){0.9}}
\put(0.2,-1.0){\vector(1,0){0.9}}
\put(1.1,-1.0){\vector(0,1){0.9}}

\end{picture}

Replacement of a link by a weighted average of links displaced
in directions perpendicular to the direction of the link amounts
to including second derivatives, $\partial^2 A_\mu / \partial_v^2$, with
$\nu\ne\mu$.  Expressing this in a gauge and Lorentz covariant form,
to lowest order in $a$ 
this modifies the action by including a term (expressed in
four-component notation, before the Kogut-Susskind spin diagonalization):
\BE
\bar\psi \left( \gamma_\mu \left[  D_\mu
+ \frac{a^2}{6} D^3_\mu
+ a^2 \frac{w}{1+6w} \left( D_\nu F_{\nu\mu}\right)
\right] + m \right) \psi
\EE
where $w$ is the staple weight.  The factor of $a^2$, required by the
dimensions of second derivative, is the expected power of the lattice
spacing for effects of lattice artifacts with staggered fermions.
(In contrast, the clover term for Wilson quarks is an order $a$
modification.)
The $D_\mu^3$ term, which violates rotational invariance, is unchanged
by the replacement of the ordinary link with a fat link. It is this term
that is canceled in the ``Naik'' (third-nearest-neighbor) derivative.

Another way to think about the use of fat links in a spectrum
computation is to consider it as the result of a modified action for the
gauge fields.  That is, we can ask what gauge action would produce links
with the probability distribution of the fat links.  In general, this
gauge action will be nonlocal, involving loops of all sizes.
Also, the fat links are not unitary matrices, so this will be an
unconventional gauge action.
For small
$w$ we could construct this gauge action as a power series in $w$.
To first order in $w$ the effect is the same as using a gauge action
consisting of the sum of the plaquettes minus $2w$ times the $2\times 1$
planar loops and the $2\times 1$ bent loops.  (The $2$ is a
combinatorial factor arising because each two-plaquette loop can be
generated in two ways, by adding a staple to either of the two
plaquettes in the loop.)  These considerations indicate that the ``fat
link'' modification of the quark action will interact with improvements
in the gauge action, and it would be dangerous to assume that the same
fattening parameter that is optimal for the Wilson gauge action will be
optimal for an improved action.

\vskip0.25in\centerline{Results}

As a dimensionless measure of flavor symmetry breaking we use the
quantity $\Delta_\pi = (m_{\pi 2} - m_\pi)/m_\rho$.  We will also use
the quantity $a m_\rho$ to define the lattice spacing, and the
dimensionless quantity $m_\pi/m_\rho$ as a measure of the quark mass.
Ideally, to compare with the
conventional quark action we would compare simulations at the same quark
mass and lattice spacing.  This requires tuning of parameters or
interpolation among various data points.

We began with a series of tests using a set of quenched lattices
with the standard Wilson gauge action at $6/g^2=5.85$\cite{MILCQUENCHED}.  
The lattice size is $20^3\times 48$.  Local meson propagators
were calculated from wall sources, using four sources in each lattice.
Because we are interested in surveying various masses and smearing
weights, only a small fraction (30 lattices) of our stored
lattices were used.
The resulting masses and mass ratios are tabulated in
Table~\ref{MASSTABLE}.
To better expose the effect of fattening the link, we have generally
chosen the same fitting range for all the values of $w$ at a given
quark mass.  (There are some exceptions where the fitting program did
not converge for the desired fit range.)
The final column of this table is the number of conjugate gradient iterations
required to converge the quark propagator calculation to a residual of 0.00005.
Table~\ref{MASSTABLE} also contains a selection of masses with the
conventional fermion action for comparison.
We also include a result at $6/g^2=6.5$ to point out that at this
lattice spacing the flavor symmetry breaking has become very
small\cite{KIMANDSINCLAIR}.

It is clear from this table that the use of fat links reduces the
flavor symmetry breaking, mostly by making the $\pi_{2}$ lighter.
As an added benefit, the smoother fat links require fewer iterations
of the conjugate gradient algorithm.
There are no obvious effects on the nucleon
to rho mass ratio.  More accurately, any change in the nucleon to
rho mass ratio is of the same order as the flavor symmetry breaking
in the rho masses, and therefore cannot be disentangled from changes
in the flavor symmetry breaking for the $\rho$'s.
It can also be seen that the improvement is quite insensitive to the
exact value of $w$.
Studies of the dependence of the flavor symmetry breaking of the
local pions on the lattice spacing with the conventional action can
be found in Refs.~\cite{SHARPEETAL,JLQCD}.
In Fig.~\ref{PIFIG} we plot the squared masses of these two pions as a
function of quark mass at this fixed lattice spacing,
and show linear extrapolations of the $\pi_2$ mass to zero quark mass.
Note that the intercept as well as the slope of the
$\pi_2$ squared masses is lower with the fat links, showing that
improvement in flavor symmetry persists in the chiral
limit $m_q \rightarrow 0$.

Because the rho and nucleon masses are reduced when $w$ is turned on,
part of the improvement in flavor symmetry breaking can be attributed to
a smaller effective lattice spacing.  To separate this effect from a
``real'' improvement, we want to compare calculations with the same
lattice spacing, which we define by the rho mass, and the same physical
quark mass, defined by $m_\pi/m_\rho$.  In looking through the available
quenched spectrum calculations with the conventional action, we find
simulations at $6/g^2=5.95$ and $6.0$ with $m_\pi/m_\rho \approx 0.65$.
We therefore chose a quark mass, 0.033, which gives a similar ratio
using fat links.  We can then compare the fat link spectrum with
$m=0.033$ and $w=0.4$ to these two conventional calculations in
the box below.  Here we see that even though
$m_\pi/m_\rho$ is slightly smaller for the fat link calculation ($\Delta_\pi$
increases as $m_q \rightarrow 0$) and the lattice spacing (from
$am_\rho$) is larger, $\Delta_\pi$ with the fat link action is about
half that for the conventional action.

\begin{center}\begin{minipage}[t]{5.0in}
\begin{center}
{\large
\begin{tabular}{|llllll|}
\hline
$6/g^2$ & $am_q$ & $w$ & $am_\rho$ & $m_\pi/m_\rho$ & $\Delta_\pi$  \\
5.85 & 0.033 & 0.40 & 0.678(11) & 0.621(10) & 0.052(3) \\
5.95 & 0.025 & 0 & 0.5954(28) & 0.651(3) & 0.107(4) \\
6.00 & 0.02 & 0 & 0.520(3) & 0.648(5) & 0.090(7) \\
\hline
\end{tabular}
} 
\end{center}
Mass ratios for the fat link action at $6/g^2=5.85$ and approximately matched
conventional calculations.  The conventional calculations are at
similar $m_\pi/m_\rho$, but at smaller lattice spacing as defined
by the $\rho$ mass.  However, the dimensionless flavor symmetry breaking
parameter is considerably smaller with the fat link action.
\end{minipage}\end{center}

\begin{table}[tph]
\setlength{\tabcolsep}{1.0mm}
\begin{tabular}{llllllllll}
\multicolumn{10}{c}{Fat link masses}\\
$w$ & $m_\pi$ & $m_{\pi 2}$ & $m_{VT}$ & $m_{PV}$ & $m_N$ &
  $m_\pi/m_\rho$ & $m_N/m_\rho$ & $\Delta_\pi$ & CG \\
\multicolumn{10}{c}{$6/g^2=5.85$, $am_q=0.01$}\\
0.00 & .273(1) & .435(15) & .60(2) & .61(3) & .88(2) & 
	.448(22) & 1.44(8) & .266(28) & 1413 \\
0.10 & .246(1) & .316(10) & .57(2) & .56(1) & .80(2) & 
	.439(14) & 1.43(4) & .125(18) & 1008 \\
0.20 & .239(1) & .292(7) & .57(2) & .56(1) & .80(2) & 
	.427(8) & 1.43(4) & .095(13) & 892 \\
0.30 & .236(1) & .290(5) & .55(1) & .56(2) & .79(2) & 
	.421(15) & 1.41(6) & .096(10) & 871 \\
0.40 & .237(1) & .288(4) & .56(2) & .56(2) & .80(2) & 
	.423(15) & 1.43(6) & .091(8) & 867 \\
\multicolumn{10}{c}{$6/g^2=5.85$, $am_q=0.02$}\\
0.00 & .379(1) & .494(8) & .675(8) & .676(14) & .998(17) & 
	.561(7) & 1.48(3) & .170(12) & 617 \\
0.10 & .342(1) & .399(3) & .617(12) & .618(12) & .903(13) & 
	.553(11) & 1.46(4) & .093(5) & 463 \\
0.20 & .332(1) & .380(3) & .607(12) & .616(14) & .885(13) & 
	.547(11) & 1.46(4) & .079(5) & 427 \\
0.30 & .330(1) & .374(3) & .604(11) & .609(13) & .880(13) & 
	.546(10) & 1.46(3) & .074(5) & 417 \\
0.40 & .329(1) & .371(3) & .603(11) & .605(13) & .878(13) & 
	.57(2) & 1.48(5) & .069(6) & 413 \\
\multicolumn{10}{c}{$6/g^2=5.85$, $am_q=0.033$}\\
0.00 & .480(1) & .599(9) & .768(20) & .781(26) & 1.11(2) & 
	.625(16) & 1.45(5) & .155(12) & 371 \\
0.10 & .437(1) & .487(2) & .699(11) & .689(15) & 1.00(2) & 
	.625(10) & 1.43(4) & .072(3) & 271 \\
0.20 & .425(1) & .466(2) & .684(11) & .684(15) & .97(2) & 
	.621(10) & 1.42(4) & .060(3) & 250 \\
0.30 & .422(1) & .459(2) & .679(11) & .682(15) & .96(2) & 
	.622(10) & 1.41(4) & .055(3) & 245 \\
0.40 & .421(1) & .456(2) & .678(11) & .682(14) & .96(2) & 
	.621(10) & 1.42(4) & .052(3) & 243 \\
0.50 & .421(1) & .455(2) & .677(10) & .683(14) & .96(2) & 
	.622(9) & 1.42(4) & .050(3) & 242 \\
\multicolumn{10}{c}{}\\
\multicolumn{10}{c}{Conventional masses}\\
$6/g^2$,$am_q$ & $m_\pi$ & $m_{\pi 2}$ & $m_{VT}$ & $m_{PV}$ & $m_N$ &
  $m_\pi/m_\rho$ & $m_N/m_\rho$ & $\Delta_\pi$ &  \\
5.85,0.02 & .3802(1) & .5075(12) & .6936(19) & ? & 1.003(5) & 
	.549(1) & 1.45(1) & .184(2) & \\
5.85,0.01 & .2736(1) & .4144(17) & .6306(42) & ? & .888(8) & 
	.434(3) & 1.41(2) & .223(3) & \\
5.95,0.025 & .3875(7) & .4512(20) & .5954(28) & ? & .893(10) & 
	.651(3) & 1.50(2) & .107(4) & \\
5.95,0.01 & .2501(9) & .3215(44) & .5159(40) & ? & .725(29) & 
	.485(4) & 1.41(6) & .138(9) & \\
6.00,0.02 & .337(2) & .384(3) & .520(3) & ? & .762(5) & 
	.648(5) & 1.46(1) & .090(7) & \\
6.00,0.01 & .241(2) & .299(9) & .481(7) & ? & .677(8) & 
	.501(8) & 1.41(3) & .121(19) & \\
6.15,0.02 & .2967(3) & .3171(9) & .426(1) & ? & .632(2) & 
	.696(2) & 1.48(3) & .048(2) & \\
6.15,0.01 & .2098(4) & .2331(8) & .372(1) & ? & .532(2) & 
	.564(2) & 1.43(7) & .063(2) & \\
6.5,0.01 & .156(2) & .157(2) & .243(4) & .246(3) & .379(3)? & 
	.642(13) & 1.56(3) & .000(12) & \\
 
\end{tabular}
\caption{
Masses and mass ratios with fat link fermion action, and comparable
spectrum results with the conventional fermion action.  The simple
plaquette gauge action was used.  All of the fat link masses were
run on the same set of configurations, so all of the masses are strongly
correlated.  The $6/g^2=6.0$ and $6.5$ masses are from
Ref.~\protect\cite{KIMANDSINCLAIR}.
\label{MASSTABLE}
}
\end{table}

\psfigure {4.0in} {1.0in} {PIFIG} {pimass.ps} {
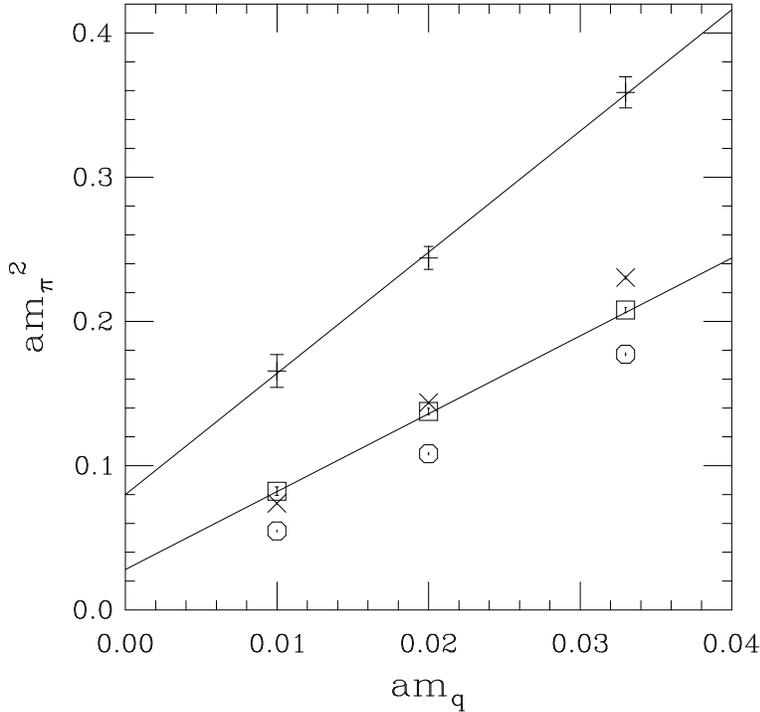
The squared pion masses versus quark mass for the Wilson gauge action at
$6/g^2=5.85$.  
The plusses and crosses are the $\pi_2$ and Goldstone $\pi$ masses with
the conventional fermion action, and the squares and octagons are the
$\pi_2$ and Goldstone $\pi$ masses with a staple weight $w=0.4$.
The lines are linear fits to the squared non-Goldstone pion masses.
}

\begin{table}[tph]
\setlength{\tabcolsep}{1.0mm}
\begin{tabular}{llllllllll}
\multicolumn{10}{c}{$6/g^2=7.40$, $am_q=0.04$, improved gauge action}\\
$w$ & $m_\pi$ & $m_{\pi 2}$ & $m_{VT}$ & $m_{PV}$ & $m_N$ &
  $m_\pi/m_\rho$ & $m_N/m_\rho$ & $\Delta_\pi$ & CG \\
0.00 & 0.5347(3) & 1.06(3) & 1.23(1) & 1.37(2) & 1.81(4) & 
	0.435(4) & 1.47(3) & 0.427(25) & 298 \\
0.10 & 0.5627(3) & 0.894(13) & 1.17(1) & 1.22(1) & 1.73(2) & 
	0.481(4) & 1.48(2) & 0.283(11) & 225 \\
0.20 & 0.5719(3) & 0.842(7) & 1.16(1) & 1.19(1) & 1.70(1) & 
	0.493(4) & 1.47(2) & 0.233(6) & 206 \\
0.30 & 0.5776(4) & 0.824(6) & 1.15(1) & 1.18(1) & 1.68(1) & 
	0.502(4) & 1.46(2) & 0.214(6) & 201 \\
0.40 & 0.5818(4) & 0.817(5) & 1.15(1) & 1.18(1) & 1.68(1) & 
	0.506(4) & 1.46(2) & 0.205(5) & 198 \\
 
\end{tabular}
\caption{
Masses and mass ratios for fat link quarks with an improved gauge action.
\label{IMPMASSTABLE}
}
\end{table}

\begin{table}[tph]
\setlength{\tabcolsep}{1.0mm}
\begin{tabular}{llllllllll}
\multicolumn{10}{c}{$6/g^2=5.85$, $am_q=0.020$, with Naik derivative}\\
$w$ & $m_\pi$ & $m_{\pi 2}$ & $m_{VT}$ & $m_{PV}$ & $m_N$ &
  $m_\pi/m_\rho$ & $m_N/m_\rho$ & $\Delta_\pi$ & CG \\
0.00 & 0.3652(7) & 0.487(8) & 0.683(20) & 0.70(4) & 0.983(17) & 
	.535(15) & 1.44(5) & 0.178(13) & 733 \\
0.10 & 0.328(1) & 0.387(3) & 0.611(13) & 0.65(5) & 0.890(14) & 
	.537(12) & 1.46(4) & 0.097(6) & 529 \\
0.20 & 0.319(1) & 0.368(3) & 0.600(13)& 0.66(5) & 0.872(14) & 
	.532(12) & 1.45(4) & 0.082(6) & 488 \\
0.30 & 0.316(1) & 0.362(3) & 0.597(13) & 0.66(4) & 0.867(14) & 
	.529(12) & 1.45(4) & 0.077(6) & 476 \\
 
\end{tabular}
\caption{
Masses and mass ratios for a fat link action including the Naik
third-nearest-neighbor term.  The gauge configurations are the same as
in table~\protect\ref{MASSTABLE}.
\label{NAIKMASSTABLE}
}
\end{table}

Motivated by the interaction of fat links and improvement of
the gauge action discussed above, as well as the fact that our
studies of quenched Kogut-Susskind spectroscopy
indicate that improvement of the gauge action results in
some reduction of flavor symmetry breaking, we calculated the fat link
spectrum on a set of stored configurations with an improved gauge
action\cite{ALFORD}.
\BE
S_g = \frac{\beta}{3}\LB 
\sum ( {\rm plaquettes}) 
- \frac{1}{20 u_0^2}\LP 1 + 0.4805 \alpha_s \RP \sum ( {\rm 2x1\ loops)}
- \frac{1}{u_0^2} 0.03325\alpha_s \sum( {\rm 1x1x1\ loops} ) \RB
\EE
Results are in table~\ref{IMPMASSTABLE}.
Again we see a dramatic improvement in the flavor symmetry breaking.
Curiously, in this case the Goldstone pion mass increases with
fattening, while it decreased in the $6/g^2=5.85$ calculation.

It is also interesting to combine the fattened links with Naik's
improvement to the fermion action.  The Naik improvement removes the
order $a^2$ (relative to the leading term) error in the quark dispersion
relation.
It simply consists of replacing the nearest neighbor term in $\Dslash$,
$U_\mu(x)\psi(x+\hat\mu)$ with a combination of nearest and third
nearest neighbor terms:
\BE  \frac{9}{8}U_\mu(x)\psi(x+\hat\mu) - \frac{1}{24}
U_\mu(x)U_\mu(x+\hat\mu) U_\mu(x+2\hat\mu)\psi(x+3\hat\mu)\ \ \ .\EE
This produces a rapid convergence of quantities like the free
field energy and baryon number susceptibility as a function of the
number of time slices, and has been used in a high temperature QCD
simulation by the Bielefeld group\cite{KARSCHLATTICE96}.  However, in
our quenched spectrum calculations\cite{MILCIMPROVED} addition of the
third neighbor term had little effect on the flavor symmetry breaking.
Since high temperature calculations are typically done with larger
lattice spacings than zero temperature calculations, it is especially
important to improve the actions here.  We may hope that a combination
of the Naik improvement with fat links or similar improvement in the
coupling of the quarks to the gauge fields could produce a simulation
with an accurate continuum free field behavior at high temperature and a
gas of the right number of light pions at low temperature.
As a first step, we calculated the spectrum at $am_q=0.02$ using
a third nearest term.  There are several decisions to be made here.
How should tadpole improvement be applied to the third neighbor term
when fat links are used?  Should the fat link in the third neighbor term
be the product of three of the single fat links, or some other weighted
combination of paths?
We began with a spectrum calculation using just the product of three
fattened links, with the unimproved coefficient $-1/24$ for the third
nearest neighbor, using the same $6/g^2=5.85$ lattices.
Results are in table~\ref{NAIKMASSTABLE}.

\vskip0.25in\centerline{Extensions and Conclusions}

We find that the replacement of the simple gauge link by a fattened link
in the fermion action reduces the flavor symmetry breaking of the
local pions by roughly a factor of two for the parameters
used here.  Since flavor symmetry breaking is expected to be
proportional to $a^2$, this corresponds to a modest increase in the
lattice spacing at which simulations of a prescribed quality can 
be carried out.  
However, the computer time required is proportional to a large power of
the lattice spacing, so a small increase in lattice spacing can
translate into a large gain in computer time.

The action considered here was primarily motivated by its simplicity and
consistency with the lattice symmetries.
It is plausible that this works because
using the fat links in the fermion action smooths
out the effects on the quarks of ultraviolet fluctuations in the gluon field,
and the flavor symmetry breaking is less severe on the
smoother configurations.
(For comments on the effects of smoothing the gluon field seen by the
quarks on the tadpole contributions, see section 3.2 in
Ref.~\cite{LEPAGE}.)
Clearly a better theoretical
understanding is wanted.  In particular, a computation of the optimum
staple weight is needed.  However, it is likely that for the relatively 
large lattice spacings at which one would like to perform simulations with 
improved actions a    
nonperturbative ({\it i.e.} empirical) determination of the coefficients will
be necessary.  It would also be very interesting to see how this action
affects rotational invariance.

We expect that an improvement of flavor symmetry in quenched spectrum
calculations is a strong indication that dynamical simulations with this
action will better reproduce the physics of a pion cloud.  This needs to
be tested.

\vskip 0.25in \centerline{Acknowledgments}
This work was supported by NSF grants
NSF--PHY96--01227, 
NSF--PHY91--16964 
and DOE contracts
DE-2FG02--91ER--40628, 
DE-AC02--86ER--40253, 
DE-FG03--95ER--40906, 
DE-FG05-85ER250000, 
DE-FG05-92ER40742, 
and 
DE-FG02--91ER--40661. 
Calculations were carried out on the Intel Paragon at the San Diego
Supercomputer Center.
We thank Tom DeGrand and Craig McNeile for useful suggestions.

\end{document}